\documentstyle[11pt]{article}
\begin{quote}
\raggedleft {NSF-ITP-93-151}
\end{quote}
\bigskip
\begin{center}
{\bf {\Large {Instantons in Schwinger model} } }
\end{center}
{\centerline {\bf A.V.Smilga} }
\begin{document}
\setlength{\baselineskip}{24.2pt}
\begin{center}
 Institute for Theoretical Physics, UCSB, Santa Barbara , CA 93106-4030,
USA\footnote {On leave of absence from ITEP, B.Cheremushkinskaya 25, Moscow,
117259, Russia.}\\

\vskip 12pt
{\bf Abstract} \\
\end{center}

The known calculations of the fermion condensate $<\bar{\psi}\psi>$ and the
correlator $<\bar{\psi}\psi(x) ~\bar{\psi}\psi(0)>$ have been interpreted in
terms of
{\em localized} instanton solutions minimizing the {\em effective} action.
Their size is of order of   massive
photon Compton wavelength $\mu^{-1}$. At high temperature, these instantons
become
quasistatic and present the 2-dimensional analog of the `walls' found recently
 in 4-dimensional gauge theories. In spite of the static nature of these
solutions, they should not be interpreted as `thermal solitons' living in
Minkowski space: the mass of these would-be solitons does not display itself
in the physical correlators.

At small but nonzero fermion mass, the high-T partition function of $QED_2$ is
saturated by the rarefied gas of instantons and antiinstantons with density
$\propto m~\exp\{-S^{inst.}\}~=~m~\exp\{-\pi T/\mu\}$ to be confronted with the
dense strongly correlated instanton-antiinstanton liquid saturating the
partition function at $T=0$.

\vskip 24 pt

\bigskip
\bigskip
\bigskip

\vfill \eject

\section{Motivation}

The appearance of instantons (topologically nontrivial Euclidean configurations
minimizing the action) is a very common feature of many different field
theories
{}~\cite{Raj}. Of particular interest are the instantons in QCD~\cite{BPTS}
which
 are so beautiful that a serious hope existed~\cite{CDG} and still exists~\cite
{Shur} that, working with instantons, one can perceive the essential features
 of QCD vacuum state.

However, soon after discovery of instantons, it has been
 understood that the {\em naive} instanton calculations meet problems. The
matter is that the quasiclassical approximation on which these calculations are
based just does not work in QCD---everything depends on the instantons of
large sizes where quantum corrections are 100\%~ important. It is impossible
 to calculate analytically these quantum corrections in QCD for $\rho \sim
\Lambda_{QCD}^{-1}$ (though for smaller $\rho$ where the quantum effects are
still
under control and can be considered as perturbation, it is possible~\cite
{vain}).

Generally, the situation is much better, however, at high temperatures $T \gg
\Lambda_{QCD}$. The high-$T$ instanton action $S^I = 8\pi^2/g^2(T)$ is large
 and quasiclassical approximation works: the amplitude of quantum fluctuations
is small compared to the amplitude of classical field. As the temperature
 grows up, the instantons cool down! That allows one to perform some explicit
instanton calculations in high-$T$ QCD which {\em are} under control. E.g., the
fermion condensate in high-$T$ QCD with {\em one} quark flavour can be found
{}~\cite{Yung}.

The {\em pure} Yang-Mills theory involves, besides instantons, also planar
topologically nontrivial Euclidean field configurations. They appear due to
nontrivial $\pi_1[{\cal G}] = Z_N$ where the true gauge group ${\cal G}$ for
the
pure Yang-Mills theory is $SU(N)/Z_N$ rather than just $SU(N)$ (gluon fields
belong to the adjoint representation and are not transormed under the action of
the elements of the center---see Ref.~\cite{bub} for the detailed discussion).
The action of these configurations involves the large area factor $\sim L^2$
where L is the size of the box, but, if the size of the box is finite (as is
always so in practical numerical calculations in QCD ), these topologically
nontrivial sectors (known as 't Hooft fluxes~\cite{flux}) do contribute in the
partition function.

At low temperatures, quasiclassical approximation does not work, and little can
be said about properties of these configurations. But at high temperatures,
quasiclassical approximation works and the characteristic field configurations
in the path integral present the classical wall-like solutions with the width
of order of Debye screening length $\sim (gT)^{-1}$ and the surface action
density\footnote{The result (~\ref{surf}) has been derived in Ref.~\cite{pisa}.
The authors of that work tried to calculate also the next-to-leading term
$\propto gT^2$ in the action density, but it has been shown in Ref.~\cite{bub1}
that this calculation is not infrared stable.  }
   \begin{equation}
   \label{surf}
   \frac{S^{SU(N)}}{{\cal A}} = \frac{4\pi^2(N-1)T^2}{3\sqrt{3N}g}
   \end{equation}
, and quantum fluctuations are relatively small.

 These high-temperature wall-like solutions turn out to be static (by the
simple reason that, when the size of the Euclidean cylinder $\beta = 1/T$
on which the theory is considered is small, higher Fourier harmonics are not
excited). Originally, they have been interpreted as real Minkowski space
domain walls separating distinct $Z_N$ high-$T$ states. However, there are
serious reasons to believe that these solutions have relevance only for the
Euclidean path integral and {\em cannot} be interpreted as real physical
objects
 in the Minkowski space~\cite{bub,bub1}. That means that the common assertion
about spontaneous breaking of $Z_N$--symmetry in high-$T$ ~Yang-Mills theory
is misleading---there is no symmetry breaking in the physical meaning of this
word as the physical domain walls separating the distinct phases do not appear.
Note that the Euclidean quasiclassical wall-like solutions exist also in
high-$T$ ~QED. Their surface action density can be found by the same token as
in nonabelian case~\cite{bub1}.

We want to emphasize that the narrow wall-like solutions appear only as the
solutions for the {\em effective} action (after adding the logarithm of
one-loop
determinant). At the pure classical level walls do not appear---the nontrivial
't Hooft fluxes still exist, but they are delocalized~\cite{flux,toron}.

\section{Schwinger model}
\setcounter{equation}{0}
Our main remark here is that the full scope of questions (of the relevance
of the instanton vacuum picture, of applicability of quasiclassical
approximation
 and of the physical meaning of the high-$T$ wall-like solutions) can be
effectively studied in the Schwinger model (2-dimensional massless QED). SM is
exactly soluble so that every reasonable question can be given an exact and
exhaustive answer and, from the other hand, resembles QCD in all gross
features. The action of the model reads
\begin{equation}
\label{SMAct}
S = \int d^2x [-\frac{1}{4}F_{\mu\nu}^2 + i\bar{\psi}{\cal D}_{\mu}\gamma_{\mu}
\psi]
\end{equation}
where ${\cal D}_{\mu} - igA_{\mu}$ and $g$ is the coupling constant having the
dimension of mass.

SM involves confinement---like in QCD, the spectrum involves only `mesons' with
the mass
\begin{equation}
\label{mu}
\mu = \frac{g}{\sqrt{\pi}}
\end{equation}
but not free fermions and photons. The axial current is anomalous:
\begin{equation}
\label{anom}
\partial_{\mu}\bar{\psi}\gamma_{\mu}\gamma^5\psi = \frac{1}{2\pi}\epsilon
_{\alpha\beta}F_{\alpha\beta}
\end{equation}
And what is most important for us here~---~it involves topologically nontrivial
Euclidean gauge field configurations with integer topological charge
\begin{equation}
\label{nu}
\nu = \frac{g}{2\pi} \int E(x)d^2x
\end{equation}
where $E = F_{01}$ ($\nu$ is the 2-dimensional analog of the Pontryagin class),
and hence the instantons---configurations with $\nu = \pm1$ realizing the
minimum of the action.

  What is the explicit form of these configurations? If proceeding along the
same lines as people usually do in QCD and looking or the configuration which
minimizes the pure {\em bosonic} Euclidean action
\begin{equation}
\label{sbos}
S_B = \frac{1}{2}\int d^2x ~E^2
\end{equation}
with the constraint $\nu = 1$, we are led to the {\em constant field strength}
solution
\[E = \frac{2\pi}{g{\cal A}}\]
where ${\cal A}$ is the total area of the Euclidean manifold on which the
theory
is defined. Introducing the compact manifold is neccessary in this approach to
provide for the infrared regularization of the theory.

Different choices for
this manifold are possible. In Ref.~\cite{indus}, the Euclidean functional
integral on the 2-dimensional sphere in different topological sectors has been
calculated. But, bearing in mind the parallels with 4-dimensional theories and
also the generalization of the results for finite temperature case, it is more
instructive to consider the theory on the 2-dimensional torus with the spatial
size $L$ and the imaginary time size $\beta$ ~\cite{Wipf}. Here we always
assume
that $L\gg\mu^{-1}$. When also $\beta\gg\mu^{-1}$, the boundary effects become
irrelevant for physical quantities
\footnote{On the large torus, the boundary effects are always exponentially
suppressed
{}~\cite{Lee,cond,LS}. On the large 2-dim sphere with $R\gg\mu^{-1}$,
they are suppressed only as a power due to finite curvature of the manifold
{}~\cite{cond}. }.
When $\beta \leq \mu^{-1}$, they are relevant and should be interpreted as the
effects due to finite temperature $T = \beta^{-1}$.

The constant field strength solution presents a fiber bundle on the torus.
Bearing in mind the subsequent discussion of the high-T case and the parallels
with 4-dim gauge theories we are going to draw, it is convenient to choose
the gauge
$A_1 = 0$. The solution then takes the form
\begin{equation}
\label{trinst}
A_0(x,\tau) = -\frac{2\pi}{gL\beta}x
\end{equation}
It satisfies the twisted boundary conditions
\begin{eqnarray}
\label{bc}
A_0(x,\beta) = A_0(x,0)
\nonumber \\
A_0(L,\tau) = A_0(0,\tau) - \frac{i}{g}\Omega^{\dagger}\partial_0\Omega
\end{eqnarray}
where
\begin{equation}
\label{Omega}
\Omega(\tau) = \exp(-2\pi i \tau/\beta)
\end{equation}
is the gauge transformation matrix. In finite temperature applications, the
corresponding b.c. for fermion fields in imaginary time directions are
antiperiodic, and the b.c. in spatial direction involve an extra gauge
transformation
\[\psi(L,\tau) = \Omega(\tau)\psi(0,\tau)\]
The transition matrix (~\ref{Omega}) satisfies self-consistency condition
$\Omega(\beta) = \Omega(0)$ which {\em is} the reason for the topological
charge (~\ref{nu}) to be quantized.

The solution (~\ref{trinst}) is the direct analog of the 4-dimensional 't Hooft
toron solutions~\cite{toron}.

The configurations with $\nu = \pm 1$ are responsible for the formation of the
fermion condensate in the Schwinger model
\begin{equation}
\label{cond}
<\bar{\psi}\psi>_{vac.} = - \frac{\mu}{2\pi}e^{\gamma}
\end{equation}
where $\gamma$ is the Euler constant. There are many ways to get this result.
The easiest way is to employ the bosonization technique~\cite{Hetrick,cond}.
But bosonization rules are specific for two dimensions, and it is more
instructive to extract the condensate from the Euclidean path integral in the
sectors with $\nu = \pm 1$. This has been done in Refs.~\cite{indus,Wipf}. In
particular, Sachs and Wipf~\cite{Wipf} considered SM on 2-dimensional torus
and calculated the functional integral over the quantum fluctuations around
the constant strength solution (~\ref{trinst}). The amplitude of these quantum
fluctuations turns out to be large (the characteristic field configurations
contributing to the path integral are rather far in Hilbert space from the
classical bosonic solution (~\ref{trinst}), and the quasiclassical {\em
picture}
does not work). The calculation is still possible, however, due to the fact
that the functional integral is {\em exactly} Gaussian.

\section{Path integrals and their saddle points.}
\setcounter{equation}{0}
Any field in the topological sector $\nu = 1$ can be decomposed as
\begin{equation}
\label{decomp}
A_\mu = A_{\mu}^{cl.i.} + A_{\mu}^{(0)} - \epsilon_{\mu\nu}\partial_\nu \phi
+ \partial_\mu \chi
\end{equation}
where $A_\mu^{(0)}$ is the constant part of the potential, $A_\mu^{cl.i.}$ is
the classical instanton solution (~\ref{trinst}), ~$\partial_\mu \chi$ is the
gauge part, and the part $-\epsilon_{\mu\nu}\partial_\nu\phi$ carries
nontrivial
dynamic information. After calculating the fermion determinant $\det\parallel
i{\cal D} - m \parallel$ ($m$ is the small fermion mass $m \ll g$), the
partition function can be written as ~\cite{Wipf}
\begin{equation}
\label{Z1}
Z_1 \propto m\int \prod d\phi \int d^2x e^{-2g\phi(x)} \exp\left\{-\frac{1}{2}
\int \phi(\Delta^2 - \mu^2\Delta)\phi d^2y\right\}
\end{equation}

Here $\phi\Delta^2\phi/2 =  (\Delta\phi)^2/2 = E^2/2$ is the classical part of
the action density, the term $\propto \mu^2\phi\Delta\phi$ in the effective
action is the
local part of the fermion determinant and gives mass to photon, and the factor
$m\int d^2x e^{-2g\phi(x)}$ comes from the fermion zero mode:~ $m$ is the
eigenvalue and the integral $\int d^2x e^{-2g\phi(x)}$ is the normalization
factor of the fermion zero mode
\begin{equation}
\label{zero}
\psi_L \propto \exp\{-g\phi(x)\}
\end{equation}

Let us find now the saddle points of the Gaussian functional integral \\
$\int\prod d\phi\exp\{-S^{eff}[\phi]\}$. To this end, we first substitute
\[ \int d^2x \exp\{-2g\phi(x)\} \rightarrow {\cal A}\exp\{-2g\phi(x_0)\} \]
where ${\cal A}$ is the total area of our Euclidean manifold [obviously, this
substitution does not change the value of the functional integral (~\ref{Z1})].
The stationary point is the solution to the equation
\begin{equation}
\label{eqmot}
(\Delta^2 - \mu^2\Delta)\phi^{eff.inst.}(x) = -2g\delta(x-x_0)
\end{equation}
Thus, $\phi^{e.i.}$ is just the Green's function of the operator
${\cal O} = \Delta^2 - \mu^2\Delta$. It has the form
\footnote{
The notion of effective (or `induced') instanton in SM has been introduced
long ago ~\cite{schroer}, but seems to be very well forgotten since that time.}
\begin{equation}
\label{inst}
\phi^{e.i.}(x) = \frac{1}{g} [ K_0(\mu|x-x_0|) + \ln(\mu|x-x_0|)
] + const
\end{equation}
[The possible linear in $x$ part of $\phi^{e.i.}(x)$ can be absorbed in the
constant component of the gauge field in the decomposition (\ref{decomp})].
The solution (~\ref{inst}) is regular at zero.
The electric field
  \begin{equation}
  \label{field}
E^{e.i.}(x) = \Delta\phi^{e.i.}(x) = \frac{g}{\pi}K_0(\mu|x-x_0|)
  \end{equation}
falls down exponentially at large distances. The topological charge (~\ref
{nu}) of this solution is equal to 1 as it should be.

We see that, in contrast to the delocalized classical instanton (~\ref{trinst})
, the effective instanton (~\ref{inst}) {\em is} localized---quantum effects
changed the properties of the solution drastically. The parameter $x_0$ should
be thought of as the collective coordinate of the center of the instanton.

Let us forget for the moment about a not fixed yet constant in the RHS in Eq.
(~\ref{inst}) and find the action of this instanton. Setting $const = 0$, we
get
\begin{equation}
\label{inac}
S^{e.i.} = g\phi^{e.i.}(0) = \ln2 - \gamma
\end{equation}
Then the {\em exact} result for the partition function (~\ref{Z1}) in the
sectors with $\nu = \pm1$ can be presented as
\begin{equation}
\label{zres}
Z_1 = Z_{-1} = m{\cal A} \frac{\mu}{2\pi}\exp\{-S^{e.i.}\}Z_0
\end{equation}

The factor $m$ comes from the fermion zero mode, the factor ${\cal A}$ arises
due to integration over collective coordinates $d^2x_0$, the factor $\mu$
appears by the dimensional reasons. The hard (in this
approach) part of the problem is to determine correctly the numerical factor
$(2\pi)^{-1}$.

To do it, one should proceed more accurately. First of all, the exact
proportionality coefficient in Eq.(~\ref{Z1}) should be found~---~more exactly,
the {\em ratio} of this coefficient to the corresponding coefficient in the
functional integral for the partition function in the topologically trivial
sector $\nu = 0$. This ratio depends on the particular way of the infrared
regularization and thereby on the size of our torus. Second, the equation
(~\ref{eqmot}) has no solution at all on a compact manifold. The correct
procedure is to project out the zero modes of the operator $\Delta^2 -
\mu^2\Delta$ by substituting $\delta(x)\rightarrow\delta(x) - 1/{\cal A}$
and imposing the constraint
\begin{equation}
\label{constr}
\int \phi(x) d^2x = 0
\end{equation}
Thereby the constant in RHS of Eq.(~\ref{inst}) is fixed.
This constant (and
hence the instanton action) also depends on the size of the torus.
\footnote{To be quite precise, the fixing of this constant requires also taking
 into account the modification of the solution to the equation (~\ref{eqmot})
 due to boundary effects which are essential when $x \sim L, \tau \sim \beta$.
 The solution has the simple form (~\ref{inst}) only at the
 vicinity of the center of the instanton and far away from the boundaries.}
For large
$L,\beta$, this dependence cancels exactly the similar dependendence of the
normalization constant, and the finite result (~\ref{zres}) is obtained
{}~\cite{Wipf}.

We cannot suggest in this respect anything new compared to the
calculation by Sachs and Wipf, but the final result (~\ref{zres}) looks so
suspiciously simple
that one is tempted to guess that an easier way to derive it may exist.
Differentiating Eq.(~\ref{zres}) over the fermion mass and adding the equal
contribution from the sector $\nu=-1$, we reproduce the result (~\ref{cond}):
\begin{equation}
\label{cond1}
<\bar{\psi}\psi> = -\frac{1}{{\cal A}Z_0} \frac{\partial}{\partial m}[Z_1
+ Z_{-1}] =
- \frac{\mu}{2\pi}e^{\gamma}
\end{equation}

There is, however, a simple {\em indirect} way to fix the coefficient in
Eq.(~\ref{cond}). To this end, one should consider the correlator $<\bar{\psi
}\psi(x)~ \bar{\psi}\psi(0)>$ in the topologically trivial sector $\nu=0$
{}~\cite
{baaq}. It can be calculated rather straightforwardly using the {\em exact}
expression for the fermion Green's function in {\em any} gauge field background
  \begin{equation}
  \label{Green}
  S_\phi(x,y) = \exp\{-g\gamma^5\phi(x)\} S_0(x-y) \exp\{-g\gamma^5\phi(y)\}
  \end{equation}
where $S_0(x-y)$ is the free fermion Green's function.
\footnote{The result (\ref{Green}) is specific for SM. Unfortunately, no
similar simple formula is known for 4-dimensional theories.}
We have
   \begin{eqnarray}
   \label{corr}
<\bar{\psi}\psi(x) ~\bar{\psi}\psi(0)>_{\nu=0}\ =\ Z_0^{-1} \int \prod d\phi
 \exp\left\{-\frac{1}{2}
\int \phi(\Delta^2 - \mu^2\Delta)\phi ~d^2y\right\} \nonumber\\
= \frac{1}{2\pi^2x^2} \int \prod d\phi  \exp\left\{-\frac{1}{2}
\int \phi(\Delta^2 - \mu^2\Delta)\phi ~d^2y\right\}
\exp\{2g[\phi(x)-\phi(0)]\}
   \end{eqnarray}
where we substituted $\cosh\{\ldots\}\rightarrow \exp\{\ldots\}$ as odd powers
of $\phi(x) - \phi(0)$ give zero after integration.

The integral is Gaussian. Its stationary point is the solution to the equation
  \begin{equation}
  \label{eqmot1}
 (\Delta^2 - \mu^2\Delta)\phi^{stat.}(y) = 2g\left[\delta(y-x) -
\delta(y)\right]
  \end{equation}
which is
   \begin{equation}
   \label{IA}
\phi^{stat.}(y) = \phi^{e.i.}(y) - \phi^{e.i.}(y-x)
   \end{equation}
with $\phi^{e.i.}(y)$ being taken from Eq.(~\ref{inst}). $\phi^{stat.}(y)$
presents an {\em instanton-antiinstanton configuration}. Note that the free
constant in the RHS of Eq.(~\ref{inst}) cancels out completely in the
difference.
\footnote
{The {\em antiinstanton-instanton configuration} $-\phi^{stat.}(y)$ plays
exactly the same role (we could substitute $\cosh\{\ldots\} \rightarrow
\exp\{-\ldots\}$ with an equal ease). A quite precise way would be to present
$\cosh\{\ldots\}$ as the sum of two exponentials and split the path integral
(\ref{corr}) in two equal parts. The stationary point of one of them
(which describes
the correlator $<\bar{\psi}_L \psi_R(x)~\bar{\psi}_R \psi_L(0)>$ ) is
$\phi^{stat.}(y)$ while the stationary point of the other corresponding to
$<\bar{\psi}_R \psi_L(x)~\bar{\psi}_L \psi_R(0)>$ is $-\phi^{stat.}(y)$.}

The calculation is standard. Introduce a new integration variable
   \begin{equation}
   \label{diff}
\varphi(y) = \phi(y) - \phi^{stat.}(y)
   \end{equation}
We have
     \begin{eqnarray}
   \label{corr1}
<\bar{\psi}\psi(x)~\bar{\psi}\psi(0)>_{\nu=0}~~ =~
{}~ Z_0^{-1} \frac{1}{2\pi^2 x^2}
\exp\{2g[\phi^{stat.}(x)-\phi^{stat.}(0)]\} \nonumber\\
  \int \prod d\varphi
 \exp\left\{-\frac{1}{2}
\int \varphi(\Delta^2 - \mu^2\Delta)\varphi ~d^2y\right\}
   \end{eqnarray}
The integral $\int \prod d\varphi \exp\{\ldots\}$ exactly cancels out the
identical functional integral for $Z_0$ and the result is
    \begin{equation}
   \label{corr2}
<\bar{\psi}\psi(x)~\bar{\psi}\psi(0)>_{\nu=0} = \frac{\mu^2}{8\pi^2}
e^{2\gamma} \exp\{-2K_0(\mu x)\}
   \end{equation}
 At large $x$, this correlator tends to a constant. One may be tempted to
extract square root of this constant and call it the fermion condensate, but
that would not be quite correct. Only the full correlator (the contributions
of all topological sectors being summed over) enjoys the cluster decomposition
property, and also the configurations with topological charge $\nu=\pm2$
contribute to the large $x$ asymptotics of the correlator. One can derive that
 \footnote
{The result (\ref{nu2}) follows from Ward identities which dictate the
particular form of $\theta$-dependence of the partition function. It holds
both in SM ~\cite{baaq} and in QCD ~\cite{LS}. In SM, it has been confirmed
 by explicit calculations ~\cite{indus,Wipf}.}
   \begin{eqnarray}
   \label{nu2}
\lim_{x \rightarrow \infty}<\bar{\psi}\psi(x) ~\bar{\psi}\psi(0)>_{\nu=2}\ =\
\lim_{x \rightarrow \infty}<\bar{\psi}\psi(x) ~\bar{\psi}\psi(0)>_{\nu=-2}
\nonumber \\ =\
\frac{1}{2}\lim_{x \rightarrow \infty}<\bar{\psi}\psi(x) ~\bar{\psi}\psi(0)>_
{\nu=0}
   \end{eqnarray}
Adding together all contributions and extracting square root, we arrive at
the result (~\ref{cond}).

   There is an important point which we want to emphasize here. In
zero-temperature SM as well as in zero-temperature QCD, the quasiclassical
picture
does not really work---the quantum fluctuations are large and the
characteristic
 field configurations in the path integral {\em have nothing to do} with either
classical (~\ref{trinst}) or effective (~\ref{inst}) instanton solutions. We
have shown in Ref.\cite{vac} that all essential properties of the
characteristic
 vacuum fields in SM are well reproduced in the model of vortex-antivortex
liquid. The basic ingredient of this model was a vortex configuration
  \begin{equation}
  \label{vort}
\phi^{vort.}(y) = \frac{1}{2g}\ln[(y-x_0)^2 + \rho^2]
  \end{equation}
It carries unit topological charge (~\ref{nu}) but does not minimize neither
classical nor effective action. The full field in our model presented a
stochastic superposition of vortices and antivortices with certain correlation
properties. All results of Ref.\cite{vac} can be, however, rederived using the
solution (~\ref{inst}) rather than (~\ref{vort}) as a basic ingredient. The
form
of the instanton is anyway distorted by quantum fluctuations and, for all
practical purposes, the configurations (~\ref{inst}) and (~\ref{vort}) are
equivalent. The really essential feature of characteristic vacuum fields is
fluctuating {\em local density} $E(x)$ of the topological charge (\ref{nu}).
Both vortex (\ref{vort}) and effective instanton (\ref{inst}) configurations
model such local fluctuations well.
We think also that all spectacular  results of the
instanton-antiinstanton liquid model in QCD ~\cite{Shur} can be reproduced
if choosing as a basic ingredient not BPST instanton but any other localized
field configuration carrying topological charge.

At large temperatures, however, the situation simplifies a lot and it is very
instructive to analyze SM in $T \gg \mu$ region.

\section{High temperatures}
\setcounter{equation}{0}
Let us repeat all derivations of the previous section for the case when the
imaginary time extension of our torus $\beta = 1/T$ is small compared to the
massive photon Compton wavelength $\mu^{-1}$ (and the spatial size is still
large $L \gg \mu^{-1}$). The partition function in the sector $\nu = 1$ is
given again by Eq.(~\ref{Z1}), only the integral $\int d^2x$ extends now over
the narrow torus. If $\beta\mu \ll 1$, one can assume the field $\phi(x,\tau)$
to be static (the contribution of higher Fourier harmonics $\propto \exp
\{2\pi ik\tau/\beta\}$ in the path integral is suppressed), and we can
substitute $\int d^2x \rightarrow \beta \int dx,\ \Delta \rightarrow \partial^2
/\partial x^2$ (from now on $x$ will always be assumed to be spatial). The
effective instanton solution satisfies the equation
  \begin{equation}
  \label{eqT}
\left( \frac{\partial^4}{\partial x^4} - \mu^2 \frac{\partial^2}{\partial x^2}
\right) \phi^{e.i.}_{high~ T}(x)~ = ~-2g\delta(x-x_0)
  \end{equation}
It has the form
  \begin{equation}
  \label{instT}
\phi^{e.i.}_{high~ T}(x) = \frac{\pi T}{g} \left[
\frac{1}{\mu}\exp\{-\mu|x-x_0|
\} + |x-x_0| \right] + const
  \end{equation}
The corresponding gauge field is
  \begin{eqnarray}
  \label{AinstT}
A_0^{e.i.}(x)\mid_{high~T} = -\frac{\partial}{\partial x} \phi^{e.i.}_{high~T}
(x) - \frac{\pi T}{g} =
\nonumber \\
\frac{\pi T}{g} \ sign(x-x_0) \left[\exp\{-\mu|x-x_0|\}
-1 \right] - \frac{\pi T}{g}
  \end{eqnarray}
(The gauge $A_1 = 0$ is chosen; the term $-\pi T/g$ comes from the properly
fixed constant term $A^{(0)}_\mu$ in the decomposition (\ref{decomp})
{}~\cite{Wipf}). When going from $x = -\infty$ to $x = \infty$,
$A_0$ changes from zero to $-2\pi T/g$ which corresponds to unit net
topological
charge
\[\nu = \frac{g\beta}{2\pi}\int E(x) dx = -\frac{g}{2\pi T}[A_0(\infty)
-A_0(-\infty)] \ =~1\]
But the electric field
  \begin{equation}
  \label{ET}
E^{e.i.}_{high~T} = -\frac{\partial}{\partial x} A_0^{e.i.}\mid_{high~T} =
T\sqrt{\pi} \exp\{-\mu|x-x_0|\}
  \end{equation}
is localized at $x \sim x_0$ in contrast to the classical solution
(~\ref{trinst}).

As we have already noticed, the notion of instanton has more physical content
at high T where quantum fluctuations are relatively small. To see this, let
us make a simple estimate ~\cite{bub1}. Let us expand the integrand in the path
integral around the classical solution (\ref{instT}--\ref{ET}) and express the
fluctuating part in terms of $a_0 = -\partial/\partial x\ [\phi(x) -
\phi^{e.i.}_{high~T}(x)]$ :
  \begin{equation}
  \label{Z0}
 Z \propto \int \prod da_0 \exp\left\{-\frac{\beta}{2}\int dx
  [(\partial_x a_0)^2 + \mu^2 a_0^2]\right\}
\end{equation}
Let us expand $a_0(x)$ in the Fourier series
\[a_0(x) = \sum_{n=-\infty}^{\infty} c_n \exp\{2\pi inx/L\},~~~~
{}~~~~~~~~~~~~~~~~~~~~c_{-n}=c_n^*    \]
 and rewrite the functional integral as the integral over $\prod_{n} dc_n$.
It is easy to see that
\[ <|c_n|^2>_{char} \ \sim\ \frac{1}{\beta L[\mu^2 + 4\pi^2n^2/L^2]}  \]
Assuming stochastic phases $\alpha_n$ for $c_{n>0}$, we may estimate
   \begin{eqnarray}
   \label{fluc}
a_0^{fluct.}(x) \ \sim\ \sqrt{LT} \sum_{n=1}^\infty \frac{\cos(\alpha_n )}
{\sqrt{4\pi^2n^2 + \mu^2 L^2}} \ \sim \nonumber \\
\sqrt{LT} \left[ \sum_{n=1}^\infty \frac{1}
{4\pi^2n^2 + \mu^2 L^2} \right]^{1/2} \ \sim \sqrt{\frac{T}{\mu}}
   \end{eqnarray}
which is much less than the amplitude of the slassical solution (\ref{AinstT})
$\sim T/\mu$. If setting $const = 0$ in the RHS of Eq.(\ref{instT}), the
action of the instanton is
   \begin{equation}
   \label{actT}
S^{e.i.}_{high~T} = g\phi^{e.i.}_{high~T}(0)~ = ~\frac{\pi T}{\mu}
   \end{equation}
The exact result for the partition function in the sectors $\nu = \pm1$
at high temperatures can be written as
   \begin{equation}
   \label{TZ1}
Z_1 = Z_{-1} = mL\exp\{-S^{e.i.}_{high~T}\}Z_0
   \end{equation}
The factor $m$ comes from the fermion zero mode and the factor $L$~---~from the
integral over the collective coordinate $x_0$ describing the spatial instanton
position. The numerical coefficient is just 1. From this, one can easily derive
   \begin{equation}
   \label{Tcond}
<\bar{\psi}\psi>_{T\gg\mu} = -\frac{1}{\beta LZ_0} \frac{\partial}{\partial m}
\left[Z_1 + Z_{-1}\right] = -2T\exp\{-\pi T/\mu\}
   \end{equation}

Again, the results (\ref{TZ1},\ \ref{Tcond}) look suspiciously simple and
again we
cannot suggest a simple direct way to derive them (other than to use the whole
machinery of the quantization in the box as in Ref.\cite{Wipf}). And again, the
simplest way to fix the exact coefficient in $<\bar{\psi}\psi>_{T\gg\mu}$ we
know of is to study the correlator $<\bar{\psi}\psi(x)~\bar{\psi}\psi(0)>_
{T\gg\mu}$ in the topologically trivial sector at large spatial $x$. The
stationary
point of the corresponding path integral is the instanton-antiinstanton
configuration
  \begin{equation}
  \label{IAT}
\phi^{stat.}_{high~T}(y) = \frac{\pi T}{g}\left[\frac{1}{\mu}\exp\{-\mu|y|\}
+ |y| - \frac{1}{\mu}\exp\{-\mu|y-x|\}
- |y-x| \right]
  \end{equation}
The integral over fluctuations cancels out with the same integral in $Z_0$, and
we get
  \begin{equation}
  \label{corrT}
<\bar{\psi}\psi(x)~\bar{\psi}\psi(0)>_
{T\gg\mu} = <\bar{\psi}\psi(x)~\bar{\psi}\psi(0)>_
{free} \exp\{2g[\phi^{stat.}_{high~T}(x) - \phi^{stat.}_{high~T}(0)]\}
  \end{equation}
where the free correlator on the cylinder is
  \begin{equation}
  \label{free}
<\bar{\psi}\psi(x)~\bar{\psi}\psi(0)>_
{T\gg\mu}\ =\ 2 T^2\ \left[\sum_{n=0}^\infty \exp\{-\pi T(2n+1)x\}\right]^2 =
\frac{T^2}{2\sinh^2(\pi Tx)}
  \end{equation}
It falls down exponentially at large distances, the exponent being given
by the lowest fermion Matsubara frequency $\omega_{min} = \pi T$.

The factor $\exp\{\ldots\}$ in the RHS of Eq.(\ref{corrT}), however, rises
exponentially at large $x$ with the same exponent so that
  \begin{equation}
  \label{asym}
\lim_{x\rightarrow\infty} <\bar{\psi}\psi(x)~\bar{\psi}\psi(0)>_
{T\gg\mu}\ = \ 2T^2\exp\{-2\pi T/\mu\}
  \end{equation}
Adding the equal contribution from the sectors $\nu = \pm2$ (cf.
Eq.(\ref{nu2}))
and taking the square root, we arrive at the result (\ref{Tcond}).

Let us look now at the solution (\ref{AinstT}) more intently. As has been
explained in details in \cite{bub}, it may be thought of as the configuration
interpolating between two adjacent minima of the high-T effective potential
$V^{eff}(A_0)$ in the constant $A_0$ background. The form of the potential is
  \begin{equation}
  \label{pot}
V^{eff}(A_0)\ =\ \frac{\mu^2}{2}\left[\left(A_0 + \frac{\pi T}{g}\right)_{mod.
{}~2\pi T/g} - \frac{\pi T}{g}\right]^2
  \end{equation}
, and it has minima exactly at $A_0^{min} = 2\pi nT/g$.

Thus, the solution (\ref{AinstT}) is much analogous to the `walls' which appear
in Euclidean path integral in 4-dimensional gauge theories
{}~\cite{pisa,bub,bub1}
--- the planar static field configurations which interpolate between different
minima of effective potential $V^{eff}(A_0)$. In Ref.\cite{pisa}, they have
been interpreted as real walls living in Minkowski space separating different
thermal $Z_N$-phases. We argued in Refs.\cite{bub,bub1} that such domain walls
do not
actually exist in Minkowski space and there is only {\em one} physical phase
both in QED and pure Yang-Mills theory at high temperature.

One of the argument comes from SM analysis---the static solutions
(\ref{AinstT})
 can (or cannot) be interpreted as real Minkowski space `solitons' with the
mass
   \begin{equation}
   \label{sol}
M^{sol.?}\ =\ TS^{e.i.}_{high T}\ =\ \frac{\pi T^2}{\mu}
   \end{equation}
exactly by the same token as the planar 4-dim Euclidean static solitons can (or
cannot) be interpreted as real walls.

Here we want to present an additional argument why they {\em can}not. If the
solitons with the mass (\ref{sol}) would really exist in some reasonable sense,
this new mass scale should have displayed itself somehow in the physical
correlators, in particular---in the correlator $<\bar{\psi}\psi(x)~\bar{\psi}
\psi(0)>$.
\footnote
{We have seen that the operator $\bar{\psi}\psi$ is `intimately connected' with
the solution (\ref{instT}--\ref{ET}). Its expectation value is proportional to
$\exp\{-S^{e.i.}_{high T}\}\ =\ \exp\{-M^{sol.?}/T\}$, and, if $\bar{\psi}\psi$
cannot be interpreted as the creation operator for this soliton, nothing can.}
But it does not. The correlator (\ref{corrT}) has three asymptotic regions.
 \begin{enumerate}
 \item At very small $x\ll T^{-1}$, it does not feel the boundaries of the box
and behaves as $\sim 1/2\pi^2x^2$.
 \item At $T^{-1}\ll x\ll\mu^{-1}$, it feels boundary but does not feel yet
fluctuating gauge fields, and is given by the expression (\ref{free}).
 \item At $x~>~\mu^{-1}$, it starts to feel the gauge field dynamics and
rapidly
levels off on a constant (\ref{asym}), the preasymptotic terms being of order
$\exp\{-\mu x\}$. The scale $\propto T^2/\mu$ is absent.
   \end{enumerate}
The full correlator
   \begin{equation}
   \label{full}
<\bar{\psi}\psi(x)\bar{\psi}\psi(0)>_T\ =\ <\bar{\psi}\psi>^{~~2}_T
\cosh\left\{
\int dk e^{ikx} \frac{1}{\sqrt{k^2+\mu^2}} \coth\frac{\sqrt{k^2+\mu^2}}{2T}
\right\}
   \end{equation}
(the contributions from all topological sectors being summed over) exhibits
the same qualitative behaviour.
\footnote
{The derivation of the result (\ref{full}) is given in the Appendix.}

\section{Partition function and characteristic field configurations at
high T.}
\setcounter{equation}{0}
As we have seen, instantons display themselves in the path integral for
$<\bar{\psi}\psi>$ and $<\bar{\psi}\psi(x)~\bar{\psi}\psi(0)>$. But it is
important to note that the path integral for the partition function itself
knows
nothing about instantons in the strictly massless Schwinger model. Really, at
$m=0$, only topologically trivial sector $\nu=0$ contributes (all
$Z_{\nu\neq0}$
involve the factor $m^{|\nu|}$ due to $|\nu|$ fermion zero modes and vanish in
the massless limit). The path integral for $Z_0$ has a trivial form (\ref{Z0})
and no instantons are seen there. To understand better what actually happens
and why instantons reappear in the path integral for the correlator
$<\bar{\psi}\psi(x)~\bar{\psi}\psi(0)>_{T \gg \mu}$ in the same topologically
trivial sector
, let us estimate the contribution of the instanton-antiinstanton pair
(\ref{IAT}) in the path integral (\ref{Z0}). The configuration $a_0^{~IA}(y)$
is depicted in Fig.1. Obviously, the corresponding contribution to $Z_0$ is
  \begin{equation}
  \label{Z0IA}
Z_0^{~IA} ~\propto~\exp\left\{-\frac{\beta x}{2}\mu^2 \left(\frac{2\pi T}{g}
\right)^2\right\}
{}~=~\exp\{-2\pi Tx\}
  \end{equation}
where $x$ is the separation between instanton and antiinstanton (we assumed
$x\gg\mu^{-1}$).

Now, the suppression (\ref{Z0IA}) can be interpreted as being
due to {\em quasizero} modes in the fermion determinant.
Really, an individual instanton (\ref{instT}) (the field configuration with
$\nu=1$) involves an exact fermion zero mode
  \begin{equation}
  \label{zeroT}
\psi_L(y)~=~ \exp\{-g\phi(y)\} ~\propto~ \exp\{-\pi T|y|\}
  \end{equation}
(where we changed the notations ~$x \rightarrow y$ and put $x_0 = 0$ ).
For instanton-antiinstanton configuration (\ref{IAT}),
the former zero mode (\ref{zeroT}) and
its counterpart for the antiinstanton located at $y=x$
\[~~~~~~~~~~~~~~~~~~~~~~~~~ \psi_R(y) = \exp\{-\pi T|y-x|\}
{}~~~~~~~~~~~~~~~~~~~~~~~~~~~~~~~~~~~~~~(5.2a)\]
are no longer exact eigenfunctions of the Dirac operator. But, if instanton
and antiinstanton are well separated, they are {\em almost} the solutions. The
true solutions and their eigenvalues can be found by solving the secular
equation taking the functions (\ref{zeroT}) and (\ref{zeroT}a) as basis and
regarding the
effects due to finite IA separation as perturbation. As a result, two
quasi-zero
modes with $\lambda\propto\exp\{-\pi Tx\}$ appear. Their product in $\det
\parallel i{\cal D} \parallel$ brings about the suppression (\ref{Z0IA}).

It is clear now why the IA configuration displayed itself in the correlator \\
$<\bar{\psi}\psi(x)~\bar{\psi}\psi(0)>$~---~the fermion operators `absorbed'
the zero modes from the fermion determinant, and the answer is finite in the
limit $x\rightarrow\infty$ (see Eq.(\ref{asym})). The finite result
(\ref{Tcond}) for the
fermion condensate $<\bar{\psi}\psi>_{T \gg \mu}$ being determined by the
path integral
in the sectors $\nu=\pm1$ is obtained exactly by the same reason.

However, the instantons {\em reappear} even in the path integral for the
partition function in high-T SM if we allow for a small but nonzero fermion
mass. It is clear why~---~the determinant factor is now $\det\parallel i
{\cal D}-m\parallel$ rather than just $\det\parallel i{\cal D}\parallel$ ,
and the contribution of the quasizero modes remains finite $\propto m^2$ even
at large $IA$ separation.

If one chooses, one may speak of the {\em confinement} of instantons in the
strictly massless case and their {\em liberation} for any small nonzero mass.
The suppression (\ref{Z0IA}) can be interpreted as being due to the linearly
rising `potential' between an instanton and antiinstanton. At $m\neq 0$, the
potential levels off at a constant at large separations. Freely moving
instantons bring about a large enthropy factor, and the contribution of the IA
configuration to $Z_0$ is
   \begin{equation}
   \label{ZIAm}
Z_0^{~IA}~=~ \left[mL\exp\{-\pi T/\mu\}\right]^2~Z_{m=0}
   \end{equation}
If the spatial box is large enough,
  \begin{equation}
  \label{kappa}
\kappa = 2mL\exp\{-\pi T/\mu\} = m|<\bar{\psi}\psi>_{T}|\beta L \gg 1
  \end{equation}
, the contribution of IA configuration to $Z_0$ {\em dominates} over purely
perturbative contribution $Z_{m=0}$ given by Eq.(\ref{Z0}).

Let us find now the contribution due to $N$ instantons and $N$ antiinstantons
(in the sector $\nu=0$, their number should be the same). It is
  \begin{equation}
  \label{ZNIA}
Z_0^{~N(IA)}~=~ \frac{1}{(N!)^2}\left[mL\exp\{-\pi T/\mu\} \right]^{2N}~Z_{m=0}
  \end{equation}
The factor $(N!)^{-2}$ appeared due to undistinguishability of instantons (and,
separately, antiinstantons). Summing over all $N$, we get
  \begin{equation}
  \label{Z0m}
Z_0 = \sum_{N=0}^\infty Z_0^{~N(IA)} =~~ Z_{m=0}~I_0(\kappa)
  \end{equation}
($I_0$ is the exponentially rising modified Bessel function). The sum
(\ref{Z0m})
is saturated at $N_{char}\sim \kappa$, i.e., when the condition (\ref{kappa})
is
satisfied, characteristic field configurations involve {\em many} (of order
$\kappa$) instanton-antiinstanton pairs.

The partition function in the topologically nontrivial sectors $\nu\neq 0$ can
be calculated in the same way. Assume for definiteness $\nu > 0$. Then the
contribution of $N+\nu$ instantons and $N$ antiinstantons in the partition
function is
  \begin{equation}
  \label{ZnuNIA}
Z_\nu\left(N_I~=~N+\nu\ ,\ N_A~=~ N\right)~=~ \frac{1}{N!(N+\nu)!}
\left(\frac{\kappa}{2}\right)^{2N+\nu}~Z_{m=0}
  \end{equation}
Summing over $N$, we get
  \begin{equation}
  \label{Znum}
Z_\nu~=~ Z_{m=0}~I_\nu(\kappa)
  \end{equation}
The full partition function is very simple (and could, of course, be written
right from the beginning from the most general premises):
    \begin{equation}
  \label{Zm}
Z~=~\sum_{\nu=-\infty}^{\infty}Z_\nu~=~Z_{m=0}\ e^{\kappa}~=~ Z_{m=0}\exp\{
-m<\bar{\psi}\psi>_T \beta L\}
  \end{equation}
The sum (\ref{Zm}) is saturated at $\nu_{char} \sim \sqrt{\kappa}$. That can
be rather easily understood. When no constraint on the net topological charge
is put, we have two independent Poisson distributions for instantons and
antiinstantons with central values of order $\kappa$. The dispersion of these
distributions and the average mean square deviation $<(N_+ -N_- )^2>^{1/2}$
are of order $\kappa^{1/2}$.

The results (\ref{Z0m},\ \ref{Znum}) are very general, they hold not only in
high-T SM, but also in QCD~\cite{LS}. They can be rigourously derived by
studying $\theta$-dependence of the partition function~\cite{LS}. It is very
instructive to see, however, how these results could be rederived here in
another way
in the case where the characteristic field configurations in the path integral
are known exactly and explicitly.

The last remark concerns topological susceptibility. It is defined as
  \begin{equation}
  \label{hidef}
\chi~=~\left(\frac{g}{2\pi}\right)^2~\int d^2x~<E(x)E(0)>~=~\frac{1}{V}<\nu^2>
  \end{equation}
where $V=L\beta$ is the total Euclidean volume. As has been mentioned,
$\nu_{char}\sim \sqrt{\kappa}$. The exact result is
  \begin{equation}
  \label{hi}
<\nu^2>~=~V\chi~=~m|<\bar{\psi}\psi>|V
  \end{equation}
It can be derived either from Ward identities \cite{Crewther} or directly
from the explicit result (\ref{Znum}) and holds universally at any temperature.
But the {\em mechanism} providing the suppression $\propto m$ in $\chi$ is
much different at high temperatures compared to that at low temperatures.
At $T \gg\mu$, the characteristic field configurations present a noninteracting
instanton-antiinstanton gas. The density of the instantons is low~---~it
involves the fermion mass factor $m$ (on top of the factor $\exp\{
-S^{e.i.}_{high~T}\} = \exp\{-\pi T/\mu\}$). This factor exhibits itself in the
topological susceptibility. From the other hand, at low temperatures, the
characteristic vacuum fields present the dense strongly interacting
instanton-antiinstanton liquid \cite{vac} (with large quantum fluctuations
distorting the shape of individual instantons). The suppression in $\chi$
appears in that case due to strong correlations in this liquid providing
the effective screening of topological charge.

It is worthwhile to mention that the whole analysis can be transferred without
essential change to QCD with {\em one} nearly massless quark. The fermion
condensate survives in this theory even at very large temperatures~\cite{Yung}
and, if the quark mass is small but nonzero and the spatial volume is large
enough, characteristic field configurations include of order
  \begin{equation}
  \label{kapQCD}
\kappa^{QCD}~~=~~m|<\bar{\psi}\psi>_T^{~QCD}|\beta V^{(3)}
  \end{equation}
instantons and about the same number of antiinstantons which do not interact
with each other (vs. dense strongly correlated instanton-antiinstanton liquid
at T=0).

\section{Acknowledgements.}

The paper was written at the Institute of Theoretical
Physics, Santa Barbara, during the research program {\em Strong Interactions at
Finite
Temperatures}. The author expresses gratitude for the warm hospitality extended
there and acknowledges the support of the National Science Foundation under
Grant No. PHY89--04035. I am indebted to E.V.Shuryak for illuminating
discussions.

\appendix
 \section{Correlator $<\bar{\psi}\psi(x)~\bar{\psi}\psi(0)>$ in
bosonization approach.}
 \setcounter{equation}{0}
Let us derive the result (\ref{full}) for the full fermion correlator
$<\bar{\psi}\psi(x)~\bar{\psi}\psi(0)>$
at finite temperature. The simplest way to do it is to use bosonization
technique~\cite{Coleman}. The lagrangian of SM is dual to the free massive
scalar field lagrangian
   \begin{equation}
  \label{Lbos}
{\cal L}_{Bos.}~=~\frac{1}{2}\left(\partial_\mu \phi\right)^2 ~-
{}~\frac{\mu^2}{2}\phi^2
  \end{equation}
and bilinear fermion combinations can be written in terms of the scalar field
$\phi$ as

\[
 \bar{\psi}\gamma_\mu\psi \equiv \frac{1}{\sqrt{\pi}}\epsilon_{\mu\nu}
\partial^\nu\phi ~~~~~~~~~~~~~~~~~~~~~~~~~~~~~~~~~~~~~~~~~~~~~~~~~~~~~~~~~~
(A.2a)\]
\[ \bar{\psi}\gamma^5\gamma_\mu\psi \equiv \frac{1}{\sqrt{\pi}}
\partial^\mu\phi~~~~~~~~~~~~~~~~~~~~~~~~~~~~~~~~~~~~~~~~~~~~~~~~~~~~~~~~~~~
(A.2b)\]
\[ \bar{\psi}\psi ~\equiv~ -\frac{\mu}{2\pi}~e^\gamma {\cal N}_\mu
\cos~\sqrt{4\pi}\phi ~~~~~~~~~~~~~~~~~~~~~~~~~~~~~~~~~~~~~~~~~~~~~~~~(A.2c)\]
\setcounter{equation}{2}
where ${\cal N}_\mu$ means normal ordering with respect to creation and
annihilation operators of the free boson field with the mass $\mu$, and the
numerical constant in Eq.(A.2c) is nothing else as the zero temperature
fermion condensate (\ref{cond}). Then all correlators of the fermion bilinears
calculated with the original SM action (\ref{SMAct}) coincide with the
correlators of the corresponding bosonized expressions calculated with the
lagrangian (\ref{Lbos}).

The typical graphs contributing to the correlator of two cosine function in
the RHS of Eq.(A.2c) are depicted in Fig.2. Note that the tadpole loops
which describe the pairing of $\phi$-operators at one and the same point and
which did not contribute at $T=0$ due to normal ordering prescription
${\cal N}_\mu$ {\em do} appear at nonzero temperature~\cite{cond}~---~an
annihilation operator gives zero when acting on the vacuum state but does
not give zero when acting on the state from thermal heat bath involving
excited states.

Working out combinatorics and employing the Gaussian nature of the path
integral, it is not difficult to see that the tadpole contributions factorize
and the result is
   \begin{equation}
  \label{corrA}
<\bar{\psi}\psi(x)~\bar{\psi}\psi(0)>_T~=~
<\bar{\psi}\psi>_T^{~~2}~\cosh\{4\pi D_T(0,x)\}
  \end{equation}
where $D_T(0,x)$ is the thermal Green's function of the scalar field and
$<\bar{\psi}\psi>_T$ has only tadpole contributions:
   \begin{equation}
  \label{condA}
<\bar{\psi}\psi>_T~=~-\frac{e^\gamma}{2\pi}\mu~\exp\left\{
-2\pi[D_T(0)-D_{T=0}(0)]\right\}
  \end{equation}
The Green's function of the Klein-Gordon operator $\Delta+\mu^2$ on a cylinder
is
   \begin{equation}
  \label{GreA}
D_T(0,x) = \frac{T}{2\pi} \sum_{n=-\infty}^\infty \int \frac{dk~e^{ikx}}
{(2\pi nT)^2 + k^2 + \mu^2} = \frac{1}{4\pi} \int dk e^{ikx} \frac{1}
{\sqrt{k^2+\mu^2}} \coth\frac{\sqrt{k^2+\mu^2}}{2T}
  \end{equation}
and we arrive at the result (\ref{full}). At $T\gg\mu$, the expression
(\ref{condA}) for $<\bar{\psi}\psi>_T$ takes the simple form (\ref{Tcond}).

\newpage
\section*{Figure Captions}

\hspace{0.25in} Fig. 1.  Instanton-antiinstanton configuration.

Fig. 2. A typical bosonized graph contributing to
$<\bar{\psi}\psi(x)~\bar{\psi}\psi(0)>_T$.


\newpage

 \begin{thebibliography}{99}
 \bibliographystyle{unsrt}

   \bibitem{Raj} R.Rajaraman, {\em Solitons and instantons: an introduction
to solitons and instantons in quantum field theory}, North Holland, New York,
1982.
   \bibitem{BPTS} A.A.Belavin, A.M.Polyakov, A.S.Schwartz, and Yu.S.Tyupkin,
Phys.Lett. {\bf B59}, 85(1975).
   \bibitem{CDG} C.G.Callan, R.Dashen, and D.J.Gross, Phys. Rev. {\bf D17},
2717
(1978); {\bf D19}, 1826 (1979).
   \bibitem{Shur} D.I.Dyakonov and V.Yu.Petrov, Nucl. Phys. {\bf B245},
259(1984);
E.V.Shuryak, Rev.Mod.Phys. {\bf 65}, 1(1993); E.V.Shuryak and
J.J.M.Verbaarschot, Nucl. Phys. {\bf B410}, 37;~55~(1993).
  \bibitem{vain} M.A.Shifman, A.I.Vainstein, and V.I.Zakharov, Nucl.Phys.
{\bf B163}, 46(1980); {\bf B165}, 45(1980);\
   A.I.Vainstein, V.I.Zakharov, V.A.Novikov, and M.A.Shifman,
Sov.Phys.Usp. {\bf 25(4)}, 196(1982).
  \bibitem{Yung} V.V.Khoze and A.V.Yung, Z.Phys. {\bf C50}, 155(1990).
  \bibitem{bub} A.V.Smilga, Bern preprint BUTP-93/3.
  \bibitem{flux} G.~ 't Hooft, Nucl.Phys. {\bf B153}, 141 (1979); Acta Physica
Austriaca Suppl. {\bf 22}, 53 (1980).
  \bibitem{pisa} T.Bhattacharaya et al., Phys.Rev.Lett. {\bf 66}, 998(1991);
Nucl.Phys. {\bf B383}, 497 (1992).
  \bibitem{bub1} A.V.Smilga, Santa Barbara preprint NSF-ITP-93-120.
  \bibitem{toron} G.~ 't Hooft, Commun.Math.Phys. {\bf 81}, 261 (1981).
  \bibitem{indus} C.Jayewardena, Helv.Phys.Acta {\bf 61}, 636(1988).
  \bibitem{Wipf} I.Sachs and A.Wipf, Helv.Phys.Acta, {\bf 65}, 652(1992).
  \bibitem{Lee} C.N.Yang and T.D.Lee, Phys.Rev. {\bf 87}, 404(1952).
  \bibitem{cond} A.V.Smilga, Phys.Lett. {\bf B278}, 371(1992).
  \bibitem{LS} H.Leutwyler and A.V.Smilga, Phys.Rev. {\bf D46}, 5607(1992).
  \bibitem{Hetrick} J.E.Hetrick and Y.Hosotani, Phys.Rev.
{\bf D38}, 2621(1988).
  \bibitem{schroer} N.K.Nielsen and B.Schroer, Phys.Lett. {\bf B66}, 373(1977).
  \bibitem{baaq} B.E.Baaquie, J.Phys. {\bf G8}, 1621(1982); E.Marinari,
G.Parisi, and C.Rebbi, Nucl.Phys. {\bf B190}, 734(1981).
  \bibitem{vac} A.V.Smilga, Phys.Rev. {\bf D46}, 5598(1992).
  \bibitem{Crewther} R.J.Crewther, Phys.Lett. {\bf B70}, 349(1977).
  \bibitem{Coleman} S.Coleman, Phys.Rev. {\bf D11}, 2088(1975).

\end{thebibliography}
\end{document}